\documentclass[aps,prd,twocolumn,superscriptaddress,nofootinbib]{revtex4-2}

\usepackage{amsmath,amssymb,graphicx,booktabs}
\usepackage{hyperref}
\usepackage{xcolor}

\hypersetup{
  hidelinks,
  pdftitle={Sign-Locked Gravitational Baryogenesis from Bulk Viscosity and Cosmological Particle Creation},
  pdfauthor={Yakov Mandel}
}

\newcommand{\mpl}{\bar M_{\rm Pl}}

\begin{document}

\title{Sign-Locked Gravitational Baryogenesis from Bulk Viscosity and Cosmological Particle Creation}

\author{Yakov Mandel}
\email{yakovm2000@gmail.com}
\affiliation{Independent Researcher, Haifa, Israel\\ORCID: 0009-0006-1766-6695}

\date{April 3, 2026}

\begin{abstract}
We study a concrete realization of gravitational baryogenesis in which a small bulk-viscous deformation of an otherwise radiation-dominated early universe generates a sign-definite curvature source.
The key point is thermodynamic irreversibility: positive entropy production makes the driving term monotonic and therefore avoids the freeze-out cancellation that suppresses rapidly oscillating or sign-changing sources.
Motivated by a simple first-order transfer-function diagnostic, we analyze the standard curvature-current operator
$\mathcal{L}_{\rm int}=(c/M^2)\,\partial_\mu R\,J^\mu_{B-L}$
in a near-radiation background with effective pressure $p_{\rm eff}=p-3\zeta H$ and $\zeta=\xi \rho/H$.
For $\xi>0$ one finds $R\neq 0$, $\dot R>0$, and a baryon asymmetry
$\eta \propto \xi T_D^5/(M^2 \mpl^3)$.
We derive the viable $(T_D,M,\xi)$ region, include entropy dilution from a finite viscous epoch, and show that the observed $\eta_{\rm obs}\simeq 8.6\times10^{-11}$ can be reproduced in a parameter region consistent with current cosmological bounds while maintaining EFT control.
The highest-scale benchmarks should be read conditionally on a very high reheating scale in view of current tensor limits.
A particle-creation sector of heavy GUT-scale fields then provides a phenomenological motivation for the required range $\xi\sim10^{-4}$--$10^{-3}$.
We also discuss the known higher-derivative instability of gravitational baryogenesis and the role of stabilized or completed embeddings.
\end{abstract}

\maketitle

\section{Introduction}
\label{sec:intro}

The observed baryon-to-entropy ratio,
\begin{equation}
\eta \equiv \frac{n_B}{s} \simeq 8.6\times10^{-11}\,,
\end{equation}
is one of the sharpest dimensionless targets for early-universe model building~\cite{Planck2018}.
Any successful scenario must satisfy the Sakharov conditions~\cite{Sakharov}: baryon-number violation, $C$ and $CP$ violation, and departure from thermal equilibrium.

A broad and economical class of models generates an effective baryon chemical potential through a derivative coupling.
In spontaneous baryogenesis one couples $(\partial_\mu\theta)J_B^\mu$~\cite{CohenKaplan87,CohenKaplan88}, while in gravitational baryogenesis one uses~\cite{Davoudiasl}
\begin{equation}
\mathcal{L}_{\rm int}=\frac{c}{M^2}\,\partial_\mu R\,J^\mu_{B-L}\,.
\label{eq:Lint_intro}
\end{equation}
Here $R$ is the Ricci scalar, $M$ is an EFT scale, and the use of $B-L$ avoids later sphaleron erasure.

Two generic obstacles appear repeatedly.
First, if the source oscillates or changes sign during freeze-out, the final asymmetry may be strongly suppressed by adiabatic averaging.
Second, in an exactly radiation-dominated universe the trace of the stress tensor vanishes, so $R=0$ and the mechanism is inactive.
The second problem can be avoided if the cosmic medium is not perfectly conformal, for example due to trace anomalies, imperfect-fluid effects, or effective creation pressures~\cite{Iorio,Antunes,Odintsov,LimaSingleton}.

The purpose of the present paper is to isolate a minimal \emph{sign-locked} realization of curvature-current baryogenesis.
We consider a short early epoch in which the radiation bath is deformed by a small bulk-viscous pressure,
\begin{equation}
p_{\rm eff}=p-3\zeta H,\qquad p=\rho/3,
\end{equation}
with $\zeta=\xi\rho/H$ and $\xi\ll1$.
This single parameter simultaneously does three things:
(i) it generates positive entropy production,
(ii) it makes the curvature trace nonzero in a near-radiation background, and
(iii) it fixes the sign of $\dot R$ during the relevant epoch.
The baryogenesis source is then monotonic rather than oscillatory.

The paper is organized as follows.
Section~\ref{sec:previous} explains how this work is positioned relative to previous literature and to our earlier entropy-clock papers.
Section~\ref{sec:transfer} introduces a compact transfer-function diagnostic for freeze-out suppression.
Section~\ref{sec:bulk} develops the bulk-viscous background and shows why $\dot R$ becomes sign-definite.
Sections~\ref{sec:baryo} and~\ref{sec:window} derive the baryon asymmetry and the viable parameter window.
Section~\ref{sec:obs} discusses entropy dilution, current cosmological constraints, and phenomenological handles.
Section~\ref{sec:micro} explains why heavy-field particle creation can plausibly motivate the required effective range of $\xi$.
Section~\ref{sec:caveat} summarizes the known instability issue and the role of stabilized or completed embeddings.

\section{Relation to Previous Work}
\label{sec:previous}

The foundational curvature-current operator of gravitational baryogenesis was introduced in Ref.~\cite{Davoudiasl}.
Later studies emphasized ways to obtain $R\neq0$ or $\dot R\neq0$ even during a nominally radiation-dominated era, including imperfect-fluid effects, particle creation, and modified-gravity constructions~\cite{Iorio,Antunes,Odintsov,LimaSingleton}.
The higher-derivative instability of the basic operator was analyzed in detail in Ref.~\cite{Arbuzova}.

The present work is also distinct from our two previous 2026 preprints~\cite{Mandel2026a,Mandel2026b}.
Ref.~\cite{Mandel2026a} introduced the transfer-function view of adiabatic freeze-out suppression and the entropy-clock idea at a general level.
Ref.~\cite{Mandel2026b} combined entropy production with a parity-violating gravitational sector.
By contrast, the aim here is narrower and more concrete:
we focus on a minimal bulk-viscous realization of the standard curvature-current operator, derive its corrected parameter window, and separate clearly the phenomenological bulk-viscous description from the microphysical particle-creation motivation.

\section{Freeze-Out Suppression as a Transfer-Function Diagnostic}
\label{sec:transfer}

Before turning to the concrete model, it is useful to state a simple diagnostic for when a time-dependent baryogenesis source survives freeze-out.

Consider the standard relaxation equation for the baryon yield $Y_B\equiv n_B/s$,
\begin{equation}
\dot Y_B + \Gamma_B\,Y_B = \Gamma_B\,Y_B^{\rm eq}(t)\,,
\label{eq:rate}
\end{equation}
where $\Gamma_B$ is the baryon-violating relaxation rate and $Y_B^{\rm eq}\propto \mu_B/T$ is the instantaneous equilibrium value induced by the source.
Near freeze-out, the relevant response time is $\tau_{\rm off}\sim \Gamma_B^{-1}$.

If the source is approximately harmonic during the last relaxation interval,
\begin{equation}
Y_B^{\rm eq}(t)=Y_\star \cos(\omega t)\,,
\end{equation}
the steady-state response of Eq.~\eqref{eq:rate} has amplitude
\begin{equation}
Y_B \sim Y_\star\,F(\omega\tau_{\rm off}),\qquad
F(x)=\frac{1}{\sqrt{1+x^2}}\,.
\label{eq:transfer}
\end{equation}
This is just the low-pass response of a first-order system.
For $\omega\tau_{\rm off}\ll1$ the source is quasi-static and survives.
For $\omega\tau_{\rm off}\gg1$ the plasma tracks only the averaged source and the final asymmetry is strongly suppressed.
A derivation is given in Appendix~\ref{app:transfer}.

\begin{figure}[t]
\centering
\includegraphics[width=\columnwidth]{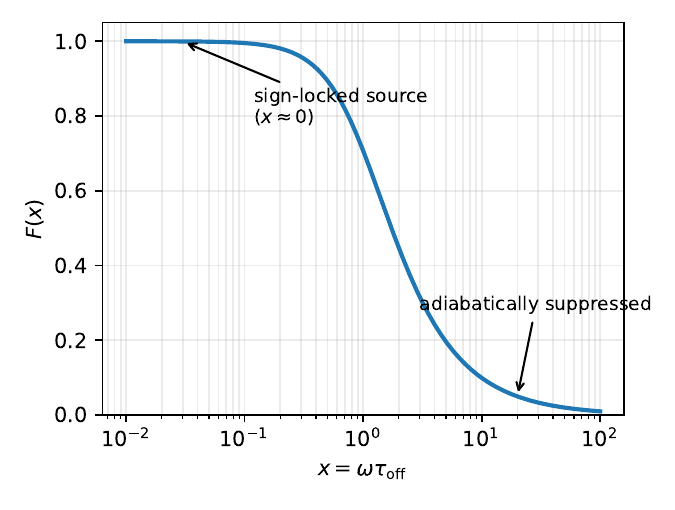}
\caption{Transfer-function diagnostic for a time-dependent baryogenesis source under smooth freeze-out.
A sign-locked source corresponds to $x=\omega\tau_{\rm off}\approx0$, while rapidly oscillating sources with $x\gg1$ are adiabatically suppressed.}
\label{fig:transfer}
\end{figure}

Equation~\eqref{eq:transfer} is not the central prediction of this paper; it is a diagnostic.
Its role here is conceptual: it explains why sign-changing sources are disfavored and motivates a source with a fixed sign throughout the relevant epoch.
The bulk-viscous realization developed below lands precisely in that regime.
Figure~\ref{fig:transfer} summarizes the diagnostic.

\section{Bulk Viscosity, Entropy Production, and a Sign-Locked Curvature Source}
\label{sec:bulk}

We work in a spatially flat FRW spacetime with $H\equiv\dot a/a$ and use natural units.
In a homogeneous background the dissipative scalar compatible with isotropy is bulk viscosity, so we write
\begin{equation}
p_{\rm eff}=p-3\zeta H,\qquad p=\frac{\rho}{3},
\label{eq:peff}
\end{equation}
and parameterize the early-time viscous strength as
\begin{equation}
\zeta=\xi\frac{\rho}{H},\qquad 0<\xi\ll1\,.
\label{eq:xi}
\end{equation}
This implies
\begin{equation}
w_{\rm eff}\equiv \frac{p_{\rm eff}}{\rho}=\frac13-3\xi,
\qquad
\varepsilon\equiv \frac{\rho-3p_{\rm eff}}{\rho}=9\xi>0.
\label{eq:weff}
\end{equation}
Thus the background remains close to radiation domination but is no longer exactly conformal.

\subsection{Entropy production}

Bulk viscosity guarantees positive entropy production,
\begin{equation}
\frac{d}{dt}(a^3 s)=\frac{9\zeta H^2}{T}\,a^3\ge0\,.
\label{eq:entropy}
\end{equation}
This monotonicity is the thermodynamic origin of sign locking in the present model.
With constant $\xi$ and approximately constant $g_*$ and $g_{*s}$, the continuity equation implies
\begin{equation}
\rho\propto a^{-(4-9\xi)},\qquad
T\propto a^{-1+9\xi/4},\qquad
a^3s\propto a^{27\xi/4}.
\label{eq:scalings}
\end{equation}

\subsection{Curvature trace and the sign of \texorpdfstring{$\dot R$}{Rdot}}

We adopt the convention
\begin{equation}
R=-8\pi G\,(\rho-3p_{\rm eff})\,.
\label{eq:Rconv}
\end{equation}
Using Eq.~\eqref{eq:weff} and $H^2=\rho/(3\mpl^2)$ gives
\begin{equation}
R=-27\xi H^2\,.
\label{eq:R}
\end{equation}
For slowly varying $\xi$,
\begin{equation}
\dot R = -54\xi H\dot H \simeq 108\,\xi\,H^3 + \mathcal{O}(\xi^2H^3)\,,
\label{eq:Rdot}
\end{equation}
where we used $\dot H=-(3/2)(1+w_{\rm eff})H^2$.
Hence
\begin{equation}
\dot R>0\qquad \text{for}\qquad \xi>0,\ H>0.
\end{equation}
The same irreversibility that makes $a^3s$ increase also fixes the sign of the curvature source.
This is the sense in which the mechanism is sign locked.

\section{Curvature-Current Baryogenesis}
\label{sec:baryo}

We now evaluate the operator
\begin{equation}
\mathcal{L}_{\rm int}=\frac{c}{M^2}\,\partial_\mu R\,J^\mu_{B-L}
\label{eq:Lint}
\end{equation}
on the background of Section~\ref{sec:bulk}.
In thermal equilibrium it induces the effective chemical potential
\begin{equation}
\mu_{B-L}=\frac{c\,\dot R}{M^2}\,.
\end{equation}
Above a decoupling temperature $T_D$, the frozen asymmetry is
\begin{equation}
\eta \equiv \frac{n_{B-L}}{s}
\simeq
\frac{15g_b}{4\pi^2 g_*}
\frac{c\,\dot R}{M^2T}\bigg|_{T_D}
\frac{1}{\Delta_S},
\label{eq:eta_pre}
\end{equation}
where $g_b$ counts the effective baryonic degrees of freedom and $\Delta_S\ge1$ accounts for any entropy growth occurring after freeze-out.

With
\begin{equation}
H(T)=\sqrt{\frac{\pi^2 g_*}{90}}\frac{T^2}{\mpl}
\end{equation}
and Eq.~\eqref{eq:Rdot}, we obtain
\begin{equation}
\boxed{
\eta \simeq
1.49\,
\frac{c\,g_b\,\xi}{\sqrt{g_*}}
\frac{T_D^5}{M^2\mpl^3}
\frac{1}{\Delta_S}
}\,.
\label{eq:eta}
\end{equation}
The structure is simple: $\xi$ measures the departure from exact conformality, $T_D^5/M^2\mpl^3$ sets the overall scale, and $\Delta_S$ keeps track of post-freeze-out dilution.

For the reference choice
\[
(\xi,T_D,M,\Delta_S)=
\left(3\times10^{-4},\,3.2\times10^{16}\ {\rm GeV},\,3.4\times10^{16}\ {\rm GeV},\,1\right),
\]
with $c\,g_b=1$ and $g_*=106.75$, Eq.~\eqref{eq:eta} reproduces $\eta_{\rm obs}$.
This already shows that the mechanism naturally lives at high scales and that the EFT requirement $T_D\lesssim M$ is nontrivial.

\section{Finite Viscous Epoch and the Parameter Window}
\label{sec:window}

We now specialize to a simple scenario:
\begin{equation}
\xi=
\begin{cases}
{\rm const}>0,& T>T_{\rm off},\\[3pt]
0,& T\le T_{\rm off}.
\end{cases}
\end{equation}
If the viscous phase continues after baryogenesis freezes out ($T_{\rm off}<T_D$), the already-produced asymmetry is diluted by the later entropy increase.
Using Eq.~\eqref{eq:scalings} one finds
\begin{equation}
\Delta_S
=
\frac{(a^3s)_{\rm off}}{(a^3s)_D}
=
\left(\frac{T_D}{T_{\rm off}}\right)^{\frac{27\xi}{4-9\xi}}
\simeq
\left(\frac{T_D}{T_{\rm off}}\right)^{27\xi/4}
\qquad (\xi\ll1).
\label{eq:DeltaS}
\end{equation}

Solving Eq.~\eqref{eq:eta} for $M$ gives
\begin{equation}
M \simeq
\left[
1.49\,\frac{c\,g_b\,\xi}{\sqrt{g_*}}
\frac{T_D^5}{\eta_{\rm obs}\mpl^3}
\frac{1}{\Delta_S}
\right]^{1/2}.
\label{eq:M}
\end{equation}
For $\Delta_S\simeq1$, a convenient scaling form is
\begin{equation}
\boxed{
\begin{aligned}
M \approx\;&
3.4\times10^{16}\ {\rm GeV}
\left(\frac{c\,g_b}{1}\right)^{1/2}
\left(\frac{106.75}{g_*}\right)^{1/4}
\\
&\times
\left(\frac{\xi}{3\times10^{-4}}\right)^{1/2}
\left(\frac{T_D}{3.2\times10^{16}\ {\rm GeV}}\right)^{5/2}
\end{aligned}
}
\label{eq:Mscale}
\end{equation}
which is the corrected version of the benchmark scaling relevant for the present paper.

\begin{figure}[t]
\centering
\includegraphics[width=\columnwidth]{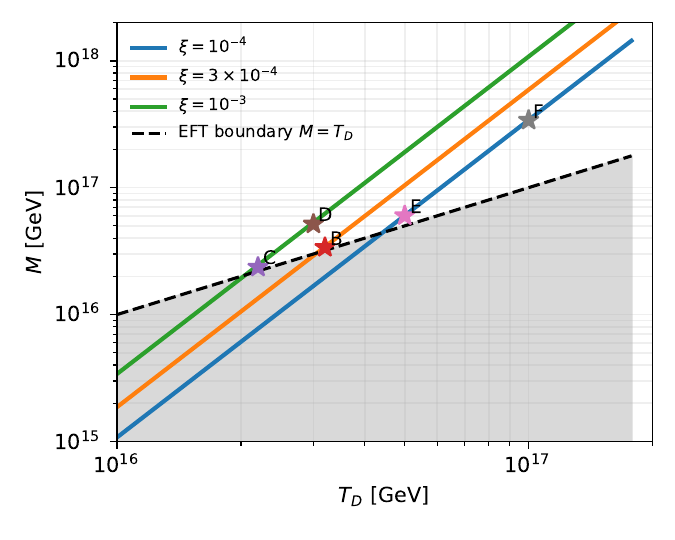}
\caption{Parameter space reproducing $\eta_{\rm obs}$ for $\Delta_S\simeq1$, $c=g_b=1$, and $g_*=106.75$.
Solid curves correspond to fixed $\xi$.
The dashed line is the EFT boundary $M=T_D$; viable points lie above it.
Stars mark the benchmark points of Table~\ref{tab:bench}, including their actual $\Delta_S$ corrections.}
\label{fig:param}
\end{figure}

Figure~\ref{fig:param} shows the parameter space that reproduces $\eta_{\rm obs}$ for $\Delta_S\simeq1$.
Contours of fixed $\xi$ are compared with the EFT boundary $M=T_D$.
The allowed region lies above the dashed line.

Table~\ref{tab:bench} lists representative benchmark points.
Points B--D illustrate finite viscous epochs with visible but still moderate entropy dilution.
Points E and F show higher-scale subcases with $\Delta_S\simeq1$ and a more comfortable EFT hierarchy.

\begin{table*}[t]
\caption{Benchmark points reproducing $\eta_{\rm obs}\simeq8.6\times10^{-11}$ for $g_*=106.75$ and $c=g_b=1$.
The ratio $M/T_D$ is shown explicitly because EFT control is one of the main consistency requirements.
The last column gives the suppression factor of any pre-existing decoupled radiation component, including primordial gravitational waves.}
\label{tab:bench}
\begin{ruledtabular}
\begin{tabular}{cccccccc}
Point & $\xi$ & $T_D$ [GeV] & $T_{\rm off}$ [GeV] & $M$ [GeV] & $M/T_D$ & $\Delta_S$ & $\Delta_S^{-4/3}$ \\
\hline
B & $3\times10^{-4}$ & $3.2\times10^{16}$ & $10^{12}$ & $3.38\times10^{16}$ & $1.06$ & $1.021$ & $0.972$ \\
C & $10^{-3}$ & $2.2\times10^{16}$ & $10^{12}$ & $2.37\times10^{16}$ & $1.08$ & $1.070$ & $0.914$ \\
D & $10^{-3}$ & $3.0\times10^{16}$ & $10^{13}$ & $5.17\times10^{16}$ & $1.72$ & $1.056$ & $0.930$ \\
E & $10^{-4}$ & $5.0\times10^{16}$ & $10^{13}$ & $6.01\times10^{16}$ & $1.20$ & $1.006$ & $0.992$ \\
F & $10^{-4}$ & $1.0\times10^{17}$ & $10^{13}$ & $3.40\times10^{17}$ & $3.40$ & $1.006$ & $0.992$ \\
\end{tabular}
\end{ruledtabular}
\end{table*}

\section{Phenomenological Handles}
\label{sec:obs}

The same entropy production that dilutes the baryon asymmetry also dilutes any decoupled relativistic component that does not share the entropy injection.
If a species $x$ has already decoupled before the end of the viscous phase, then
\begin{equation}
\left.\frac{\rho_x}{\rho_\gamma}\right|_{\rm after}
=
\left.\frac{\rho_x}{\rho_\gamma}\right|_{\rm before}
\Delta_S^{-4/3}.
\label{eq:gw}
\end{equation}
A primordial inflationary gravitational-wave background is the most obvious example.
The effect is modest for $\xi\sim10^{-4}$ but can reach the $\mathcal{O}(10\%)$ level for $\xi\sim10^{-3}$ combined with a sufficiently long finite viscous epoch, as illustrated by point C in Table~\ref{tab:bench}.

\begin{figure*}[t]
\centering
\includegraphics[width=\textwidth]{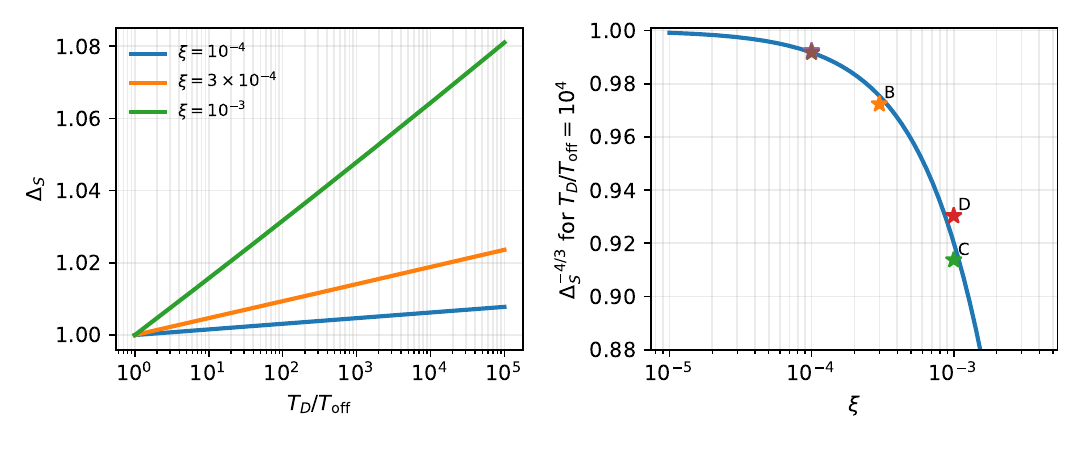}
\caption{\textit{Left:} entropy dilution factor $\Delta_S$ as a function of the duration of the viscous phase, expressed through $T_D/T_{\rm off}$.
\textit{Right:} suppression factor $\Delta_S^{-4/3}$ for a decoupled radiation component as a function of $\xi$ for the illustrative choice $T_D/T_{\rm off}=10^4$.
Benchmark points B--D are marked for orientation.}
\label{fig:obs}
\end{figure*}

Figure~\ref{fig:obs} shows the entropy-dilution factor and the associated gravitational-wave suppression.
The main lesson is simple:
for the parameter range relevant to baryogenesis, the model predicts either negligible or mildly percent-level dilution unless the viscous phase extends over many decades after freeze-out.

A second, less direct handle is the modified early-time equation of state,
\begin{equation}
w_{\rm eff}=\frac13-3\xi,
\end{equation}
which slightly changes the pre-BBN expansion history while the viscous phase is active.
For the values considered here this is a small effect, but it provides a consistency relation:
$\eta$, $\Delta_S$, and the duration of the viscous phase are not independent.

\subsection{Comparison with Current Cosmological Constraints}
\label{sec:data}

\paragraph{Baryon-abundance target.}
The normalization used in this paper, $\eta_{\rm obs}\simeq 8.6\times10^{-11}$, remains consistent with the standard CMB determination and with recent BBN reassessments of the baryon density~\cite{Planck2018,Schoeneberg2024}.
At the level of accuracy relevant for the present background-level analysis, these updates do not materially shift the baryogenesis window derived from Eq.~\eqref{eq:eta}.
The role of the observed asymmetry is therefore to set the target normalization, not to select among the benchmark microphysical realizations considered here.

\paragraph{Effective number of relativistic species.}
Recent analyses of ACT DR6 data and of joint BBN+CMB+BAO data find no evidence for excess dark radiation and give values of $N_{\rm eff}$ close to the standard prediction~\cite{ACTDR6,Goldstein2026}.
For that reason, the dilution factor $\Delta_S^{-4/3}$ in Eq.~\eqref{eq:gw} should be interpreted here as a consistency condition on any decoupled radiation component, not as an explanation of an established anomaly.
In the benchmark region of Table~\ref{tab:bench}, the predicted effect is negligible to mildly percent level unless the viscous phase persists for many decades after freeze-out.

\paragraph{Tensor bound and reheating scale.}
The current published BICEP/Keck bound, $r_{0.05}<0.036$ at $95\%$ C.L., continues to limit the inflationary energy scale~\cite{BICEPKeck2024}.
Our mechanism does not require a tensor detection, but the highest-scale benchmarks with $T_D\sim10^{16}$--$10^{17}$~GeV should be read conditionally on a very high reheating scale, close to the upper end usually considered plausible under standard inflationary assumptions.
Accordingly, the most conservative realizations of the present mechanism are the ones with the mildest hierarchy, $T_D\lesssim M$, and with only modest post-freeze-out entropy production.

\section{Particle Creation as a Microphysical Motivation for \texorpdfstring{$\xi$}{xi}}
\label{sec:micro}

So far $\xi$ has been treated as an effective parameter.
A natural question is whether the required magnitude can be motivated by plausible early-universe microphysics.

A useful clue comes from open-system cosmology and the particle-creation literature.
At the homogeneous background level, gravitational particle production can be encoded by an effective negative creation pressure, and several works have emphasized its close phenomenological relation to bulk pressure~\cite{LimaSingleton,Parker,Zeldovich,HuParker,Calzetta}.
At the same time, the two descriptions are not strictly identical from a full thermodynamic standpoint~\cite{LimaSingleton}.
For that reason we do \emph{not} claim a one-to-one microscopic equivalence.
Instead, we use particle creation as a motivation for the size and duration of the effective bulk-viscous deformation.

A conservative way to parametrize the effect of a heavy sector with multiplicity $N_h$ and characteristic mass $m$ is
\begin{equation}
\zeta_{\rm eff}^{\rm (pc)} \sim \alpha\, N_h\, m^2 H,
\label{eq:zetaeff}
\end{equation}
where $\alpha$ is a dimensionless efficiency factor encoding the details of the non-equilibrium production process.
Using $\rho=3\mpl^2H^2$, this becomes
\begin{equation}
\xi_{\rm eff}^{\rm (pc)}
\equiv
\frac{\zeta_{\rm eff}^{\rm (pc)} H}{\rho}
\sim
\frac{\alpha N_h}{3}\frac{m^2}{\mpl^2}.
\label{eq:xieff}
\end{equation}
Numerically,
\begin{equation}
\xi_{\rm eff}^{\rm (pc)}
\sim
1.7\times10^{-4}
\left(\frac{\alpha}{10^{-2}}\right)
\left(\frac{N_h}{100}\right)
\left(\frac{m}{3\times10^{16}\ {\rm GeV}}\right)^2.
\label{eq:xieff_num}
\end{equation}
This lands naturally in the range relevant for the baryogenesis window found above.

The same picture also suggests a natural shut-off scale.
Once the temperature drops below the heavy threshold, particle production becomes inefficient and the effective deformation should decay rapidly.
In the most economical subcase one expects $T_{\rm off}\sim m$ and often also $T_D\lesssim m$, which favors benchmarks with $\Delta_S\simeq1$.
The more strongly diluted points in Table~\ref{tab:bench} should therefore be viewed as phenomenological finite-duration extensions of this minimal microphysical picture rather than as the most literal threshold realizations.

\section{Caveat: Higher-Derivative Instability and Stabilized Completions}
\label{sec:caveat}

A standard caveat of gravitational baryogenesis is that if the operator in Eq.~\eqref{eq:Lint} is inserted fully self-consistently into the gravitational field equations, the dynamics for $R$ can become higher-derivative and generically unstable in part of parameter space~\cite{Arbuzova}.
The present paper does not solve that issue.
Our treatment is deliberately phenomenological:
we compute the background using Einstein gravity plus an effective imperfect fluid, and then evaluate the induced chemical potential on that background.

This issue has become sharper rather than weaker in the recent literature.
Ref.~\cite{Arbuzova} showed explicitly that the original operator generically induces a fourth-order instability and discussed stabilization by an added $R^2$ term, albeit with non-negligible cosmological backreaction.
Ref.~\cite{Pereira2024} proposed a scalar-tensor completion in which the relevant CP-violating bias is carried by scalar degrees of freedom associated with modified gravity.
Most recently, Ref.~\cite{Pereira2026} revisited the operator beyond the spectator approximation and derived the action-level backreaction and realization-dependent corrections to the effective gravitational dynamics.

For the present paper this means the following.
The bulk-viscous background developed above should be interpreted as a background-level parametric window that any stabilized or completed realization must reproduce in the controlled limit.
The approximation used here is most defensible when the departure from radiation domination is small ($\xi\ll1$), the asymmetry is tiny, and the EFT hierarchy $T_D\lesssim M$ is respected.
A fully satisfactory microscopic theory should still explain both the origin of the effective viscous deformation and the stability of the curvature sector at action level.

\section{Discussion and Conclusions}
\label{sec:conclusions}

The point of this paper is not to introduce a new operator, but to identify a minimal background in which the standard curvature-current operator becomes both active and sign definite.

The logic is compact.

First, a time-dependent baryogenesis source is filtered by freeze-out.
The transfer-function diagnostic of Section~\ref{sec:transfer} explains why rapidly oscillating or alternating-sign sources are vulnerable to strong suppression.

Second, a small bulk-viscous deformation of a radiation bath automatically produces entropy and therefore defines a monotonic arrow of time.
That same deformation generates $R\neq0$ and, more importantly, $\dot R>0$.
The source is therefore sign locked.

Third, once the source is inserted into the usual gravitational-baryogenesis operator, the observed baryon asymmetry fixes a narrow combination of scales.
After enforcing the EFT condition $T_D\lesssim M$, the preferred region is shifted upward relative to a naive low-scale estimate.
In particular, viable benchmark points live naturally around
\begin{equation}
\xi\sim10^{-4}\text{--}10^{-3},\qquad
T_D\sim10^{16}\text{--}10^{17}\ {\rm GeV},\qquad
M\sim10^{16}\text{--}10^{18}\ {\rm GeV},
\end{equation}
with only mild entropy dilution in the particle-creation-motivated subcase.

Finally, heavy-field particle creation provides a plausible phenomenological motivation for the required effective deformation.
It should be understood as motivation, not as a strict derivation of microscopic equivalence.
That distinction matters.

Relative to current data, the framework is best viewed as cosmologically allowed rather than observationally selected.
The baryon target remains standard, the induced dilution of decoupled radiation must stay compatible with current $N_{\rm eff}$ bounds, and the highest-scale benchmarks are conditional on a very high reheating scale in light of present tensor limits.

In short, the framework developed here turns the general entropy-clock intuition into a separate concrete paper:
a bulk-viscous, sign-locked realization of curvature-current baryogenesis with a corrected parameter window, explicit consistency conditions, compatibility with current cosmological bounds, and a clear distinction between effective bulk pressure and its particle-creation motivation.

\begin{acknowledgments}
The author thanks the readers of earlier drafts for useful comments.
\end{acknowledgments}

\appendix

\section{Transfer function from a driven first-order response}
\label{app:transfer}

For completeness, we derive Eq.~\eqref{eq:transfer} in the simplest local approximation.
Take Eq.~\eqref{eq:rate} with constant $\Gamma_B$ during the last relaxation interval and a harmonic source,
\begin{equation}
Y_B^{\rm eq}(t)=Y_\star \cos(\omega t).
\end{equation}
Using a complex source $Y_\star e^{i\omega t}$, the particular solution is
\begin{equation}
Y_B(t)=\frac{\Gamma_B}{\Gamma_B+i\omega}\,Y_\star e^{i\omega t}.
\end{equation}
Hence the response amplitude is
\begin{equation}
|Y_B| = Y_\star \frac{\Gamma_B}{\sqrt{\Gamma_B^2+\omega^2}}
=Y_\star \frac{1}{\sqrt{1+(\omega/\Gamma_B)^2}}.
\end{equation}
Identifying $\tau_{\rm off}\sim\Gamma_B^{-1}$ gives
\begin{equation}
F(\omega\tau_{\rm off})=\frac{1}{\sqrt{1+(\omega\tau_{\rm off})^2}},
\end{equation}
which is the result quoted in the main text.
The response also carries a phase lag $\phi=\arctan(\omega/\Gamma_B)$, but only the amplitude matters for the suppression estimate used here.

\section{Entropy dilution formula}
\label{app:dilution}

Equation~\eqref{eq:scalings} implies
\begin{equation}
a^3s\propto a^{27\xi/4},
\qquad
T\propto a^{-1+9\xi/4}.
\end{equation}
Eliminating $a$ gives
\begin{equation}
a^3s \propto T^{-\,27\xi/(4-9\xi)}.
\end{equation}
Therefore, if the viscous phase lasts from freeze-out at $T_D$ down to $T_{\rm off}$,
\begin{equation}
\Delta_S
=
\frac{(a^3s)_{\rm off}}{(a^3s)_D}
=
\left(\frac{T_D}{T_{\rm off}}\right)^{27\xi/(4-9\xi)}.
\end{equation}
For $\xi\ll1$ this reduces to the approximate expression used for quick estimates,
\begin{equation}
\Delta_S\simeq \left(\frac{T_D}{T_{\rm off}}\right)^{27\xi/4}.
\end{equation}


\end{document}